# Mining Statistically Significant Substrings using the Chi-Square Statistic

Mayank Sachan
mayasac@cse.iitk.ac.in

Arnab Bhattacharya
arnabb@iitk.ac.in

Dept. of Computer Science and Engineering
Indian Institute of Technology, Kanpur
INDIA

## ABSTRACT

The problem of identification of statistically significant patterns in a sequence of data has been applied to many domains such as intrusion detection systems, financial models, web-click records, automated monitoring systems, computational biology, cryptology, and text analysis. An observed pattern of events is deemed to be statistically significant if it is unlikely to have occurred due to randomness or chance alone. We use the *chi-square statistic* as a quantitative measure of statistical significance. Given a string of characters generated from a memoryless Bernoulli model, the problem is to identify the substring for which the empirical distribution of single letters deviates the most from the distribution expected from the generative Bernoulli model. This deviation is captured using the chi-square measure. The most significant substring (MSS) of a string is thus defined as the substring having the highest chi-square value. Till date, to the best of our knowledge, there does not exist any algorithm to find the MSS in better than $O(n^2)$ time, where $n$ denotes the length of the string. In this paper, we propose an algorithm to find the most significant substring, whose running time is $O(n^{3/2})$ with high probability. We also study some variants of this problem such as finding the top-t set, finding all substrings having chi-square greater than a fixed threshold and finding the MSS among substrings greater than a given length. We experimentally demonstrate the asymptotic behavior of the MSS on varying the string size and alphabet size. We also describe some applications of our algorithm on cryptology and real world data from finance and sports. Finally, we compare our technique with the existing heuristics for finding the MSS.

## 1. MOTIVATION

Statistical significance is used to ascertain whether the outcome of a given experiment can be ascribed to some extraneous factors or is solely due to chance. Given a string composed of characters from an alphabet $\Sigma = \{a_1, a_2, \ldots, a_k\}$ of constant size $k$, the null hypothesis assumes that the letters of the string are generated from a *memoryless Bernoulli model*. Each letter of the string is drawn randomly and independently from a fixed multinomial probability distribution $P = \{p_1, p_2, \ldots, p_k\}$ where $p_i$ denotes the probability of occurrence of character $a_i$ in the alphabet ($\sum p_i = 1$). The objective is to find the connected subregion of the string (i.e., a substring) for which the empirical distribution of single letters deviates the most from the distribution given by the Bernoulli model.

Detection of statistically relevant patterns in a sequence of events has drawn significant interest in the computer science community and has been diversely applied in many fields including molecular biology, cryptology, telecommunications, intrusion detection, automated monitoring, text mining, and financial modeling. The applications in computational biology include assessing the over representation of exceptional patterns [7] and studying the mutation characteristics in the protein sequence of an organism by identifying the sudden changes in their mutation rates [18]. Different studies suggest detecting intrusions in various information systems by searching for hidden patterns that are unlikely to occur [26, 27]. In telecommunication, it has been applied to detect periods of heavy traffic [13]. It has also been used in analyzing financial time series to reveal hidden temporal patterns that are characteristic and predictive of time series events [22] and to predict stock prices [17].

Quantifying a substring as *statistically significant* depends on the statistical model used to calculate the deviation of the empirical distribution of single letters from its expected nature. The exact formulation of statistical significance depends on the metric used; *p-value* and *z-score* [23, 25] represent the two most commonly used ones (some of the other ones are reviewed in [10, 24]). Research indicates that in most practical cases, p-value provides more precise and accurate results as compared to z-score [7].

The *p-value* is defined as the probability of obtaining a test statistic at least as extreme as the one that was actually observed assuming the null hypothesis to be true. For example, in an experiment to determine whether a coin is fair, suppose it turns up head on 19 out of 20 tosses. Assuming the null hypothesis, i.e., the coin is fair, to be true, the p-value is equal to the probability of observing 19 or more heads in 20 flips of a fair coin:[1]

$$\text{p-value} = Pr(19H) + Pr(20H) = \frac{\binom{20}{19} + \binom{20}{20}}{2^{20}} \approx 0.002\%$$

Traditionally, the decision to reject or fail to reject the null hypothesis is based on a pre-defined significance level $\alpha$. If the p-value is low, the result is less likely assuming the null hypothesis to be true. Consequently, the observation is statistically more significant.

---
[1] This definition of p-value is part of a one-sided test; however, we can also calculate the probability of getting at least 19 heads or at least 19 tails which is part of a two-sided test. The p-value is just double in this case due to symmetry.

In a *memoryless Bernoulli multinomial model*, the probability of observing a configuration $\beta_0$, given by a count vector $C = \{Y_1, Y_2, \ldots, Y_k\}$ with $\sum_{i=1}^{k} Y_i = l$ (where $l$ is the length of the substring) denoting the set of observed frequencies of each character in the alphabet, is defined as

$$Pr(C = \beta_0) = l! \prod_{i=1}^{k} \frac{p_i^{Y_i}}{Y_i!} \quad (1)$$

The p-value for this model then is

$$\text{p-value} = \sum_{\beta \text{ more extreme than } \beta_0} Pr(\beta) \quad (2)$$

However, computing the p-value exactly requires analyzing all possible outcomes of the experiment which are potentially exponential in number, thereby rendering the computation impractical. Moreover, it has been shown that for large samples, asymptotic approximations are accurate enough and easier to calculate [24].

The two broadly used approximations are the *likelihood ratio statistic* and the *Pearson's chi-square statistic* [24]. In case of likelihood ratio test, an alternative hypothesis is set up under which each $p_i$ is replaced by its maximum likelihood estimate $\pi_i = x_i/n$ with the exact probability of a configuration under null hypothesis defined similarly as in the previous case. The natural logarithm of the ratio between these two probabilities multiplied by $-2$ is then the statistic for the likelihood ratio test:

$$-2\ln(LR) = -2 \sum_{i=1}^{k} x_i \ln\left(\frac{\pi_i}{p_i}\right) \quad (3)$$

Alternatively, the Pearson's chi-square statistic, denoted by $X^2$, measures the deviation of observed frequency distribution from the theoretical distribution [5]:

$$X^2 = \sum_{i=1}^{k} \frac{(O_i - E_i)^2}{E_i} = \sum_{i=1}^{k} \frac{(Y_i - lp_i)^2}{lp_i} \quad (4)$$

where $O_i$ and $E_i$ are theoretical and observed frequencies of the characters in the substring. Since each letter of the substring is drawn from a fixed probability distribution, the expected frequency $E_i$ of a character in the substring is obtained by multiplying the length of the substring $l$ with the probability of occurrence of that character. Hence, the expected frequency vector is given by $E = lP$, where $P = \{p_1, p_2, \ldots, p_k\}$. The chi-square ($X^2$) definition in (4) can be further simplified as:

$$X^2 = \sum_{i=1}^{k} \frac{(Y_i - lp_i)^2}{lp_i} = \sum_{i=1}^{k} \frac{Y_i^2}{lp_i} - 2\sum_{i=1}^{k} Y_i + l \sum_{i=1}^{k} p_i$$

$$= \sum_{i=1}^{k} \frac{Y_i^2}{lp_i} - l \quad \left[ \because \sum_{i=1}^{k} Y_i = l \text{ and } \sum_{i=1}^{k} p_i = 1 \right] \quad (5)$$

Note that the chi-square value for a substring depends only on the count of the characters in it, and *not* on the order in which they appear. It can be seen in the coin toss example that all the outcomes that are less likely to occur have higher $X^2$ values than the observed outcome. For multinomial models, under the null hypothesis, both $X^2$ statistic and $-2\ln(LR)$ statistic converge to the $\chi^2$ distribution with $k-1$ degrees of freedom [21, 24]. Hence, the p-value of the outcome can then be computed using the cumulative distribution function (cdf) $F(x)$ of the $\chi^2(k-1)$ distribution. If $z_0$ is the $X^2$ value of the observed outcome, then its p-value is $1 - F(z_0)$.

Moreover, it has also been shown that the $X^2$ statistic converges to the $\chi^2$ distribution from below as opposed to the $-2\ln(LR)$ statistic which converges from above [21, 24]. Thus, the chi-square statistic diminishes the probability of type-I errors (false positives). Considering these significant advantages, we adopt the Pearson's $X^2$ statistic as the estimate to quantify the statistical significance in our study.

In this paper, we focus on the problem where only portions of the string instead of the whole string may deviate from the expected behavior. As discussed in the experimental section, this problem is particularly useful in the analysis of temporal strings where an external event occurring in the middle of a string may be causing the particular substring to deviate significantly from the expected behavior by inflating or deflating the probabilities of occurrence of some characters in the alphabet. Our work focuses on the problem of identification of such statistically significant substrings in large strings. Before venturing forward, we formally define the different problem statements handled in this paper for a string $S$ of length $n$.

PROBLEM 1 (MOST SIGNIFICANT SUBSTRING). *Find the most significant substring (MSS) of $S$, which is the substring having the highest chi-square value ($X^2$) among all possible substrings.*

PROBLEM 2 (TOP-T SUBSTRINGS). *Find the top-t set $T$ of $t$ substrings such that $|T| = t$ and for any two arbitrary substrings $S_1 \in T$ and $S_2 \notin T$, $X_{S_1}^2 \geq X_{S_2}^2$.*

PROBLEM 3 (SIGNIFICANCE GREATER THAN THRESHOLD). *Find all substrings having chi-square value ($X^2$) greater than a given threshold $\alpha_0$.*

PROBLEM 4 (MSS GREATER THAN GIVEN LENGTH). *Find the substring having the highest chi-square value ($X^2$) among all substrings of length greater than $\gamma_0$.*

The rest of the paper is organized as follows. Section 2 provides an overview of the related work. Section 3 formulates some important definitions and observations used by our algorithm. Section 4 describes the algorithm for finding the MSS of a string. Section 5 presents the analysis of the algorithm. Section 6 extends the MSS finding algorithm to the more general problems. Section 7 shows the experimental analysis and some applications of the algorithm on real datasets. Finally, Section 8 discusses possible future work.

## 2. RELATED WORK

The problem of identifying frequent and statistically relevant subsequences (not necessarily contiguous) in a sequence has been an active area of research over the past decade [19]. The problem of finding statistically significant subsequences within a window of size $w$ has also been addressed [3, 15]. Since the number of subsequences grows exponentially with $w$, the task of computing subsequences within a large window is practically infeasible.

We address a different version of the problem where the window size can be arbitrarily large but statistically significant patterns are constrained to be contiguous, thus forming substrings of the given string. The problem has many relevant applications in places where the extraneous factor that triggers such unexpected patterns occur continuously over an arbitrarily large period in the course of a sequence, as in the case of temporal strings. As the possible number of substrings reduces to $O(n^2)$, the problem of computing statistically significant patterns becomes much more scalable. However, it is still computationally intensive for large data.

The trivial algorithm proceeds by checking all $O(n^2)$ possible substrings. Some improvements such as blocking technique and heap strategy were proposed, but they showed no asymptotic improvement in the time complexity [2]. Two algorithms, namely,

ARLM and AGMM, were proposed which use local maxima to find the MSS [9]. It was claimed (only through a conjecture and not a proof) that ARLM would find the MSS. However, the time complexity is still $O(n^2)$ with only constant time improvements. AGMM was a $O(n)$ time heuristic that found a substring whose $X^2$ value was roughly close to the $X^2$ value of MSS, but no theoretical guarantees were provided on the bound of the approximation ratio. The comparative analysis of our algorithms with them is shown in detail in Section 7. To the best of our knowledge, no algorithm exists till date that exactly finds the MSS or solves the other variants of the problem in better than $O(n^2)$ time.

It may seem that a fast algorithm can be obtained using the suffix tree[2] [14]. However, the problem at hand is different. To compute the $X^2$ value of any substring we need not traverse the whole substring; rather, we just need the number of occurrences of each character in that substring. This can be easily computed in $O(1)$ time by maintaining $k$ count arrays, one for each character of the alphabet, where $i^{th}$ element of the array stores the number of occurrences of the character till $i^{th}$ position in the string. Each array can be preprocessed in $O(n)$ time. Furthermore, due to complex non-linear nature of the $X^2$ function we assume that no obvious properties of the suffix trees or its invariants can be utilized.

The trivial algorithm checks for all possible substrings that have $O(n)$ starting positions and for each starting position have $O(n)$ ending positions, thus requiring $O(n^2)$ time. Our algorithm also considers all the $O(n)$ starting positions, but for a particular starting position, it does not check all possible ending positions. Rather, it skips ending positions that cannot generate candidates for the MSS or the *top-t* set. We show that for a particular starting position, we check only $O(\sqrt{n})$ different ending positions, thereby scanning a total of only $O(n^{3/2})$ substrings. We formally show that the running time of our algorithm is $O(n^{3/2})$. We also extend the algorithm for finding the *top-t* substrings and other variants, all of which, again, run in $O(n^{3/2})$ time.

## 3. DEFINITIONS AND OBSERVATIONS

In the rest of the paper, any string $S$ over a multinomial alphabet $\Sigma = \{a_1, a_2, \ldots, a_k\}$ and drawn from a fixed probability distribution $P = \{p_1, p_2, \ldots, p_k\}$ is phrased as "$S$ over $(\Sigma, P)$". For a given string $S$ of length $n$, $S[i]$ ($1 \leq i \leq n$) denotes the $i^{th}$ letter of the string $S$ and $S[i \ldots j]$ denotes the substring of $S$ from index $i$ to index $j$, both included. So, the complete string $S$ can also be denoted by $S[1 \ldots n]$.

DEFINITION 1 (CHAIN COVER). *For any string $S$ of length $l$, a string $\lambda(S, a_i, l_1)$ of length $l + l_1$ is said to be the* chain cover *of $S$ over $l_1$ symbols of character $a_i$ if $S$ is the prefix of $\lambda(S, a_i, l_1)$ and the last $l_1$ positions of $\lambda(S, a_i, l_1)$ are occupied by the character $a_i$. Alternatively, $\lambda(S, a_i, l_1)$ is of the form $S$ followed by $l_1$ occurrences of character $a_i$.*

For example, if $S = cdcbbc$ then $\lambda(S, d, 3) = cdcbbcddd$, and if $S = baacd$ then $\lambda(S, a, 2) = baacdaa$.

We first prove that for any string $S$ of length $l$, $X^2$ value of any string $S'$ of length less than or equal to $l + l_1$ and having $S$ as its prefix is upper bounded by the $X^2$ value of a chain cover of $S$ over $l_1$ symbols of some character $a_i \in \Sigma$.

---

[2]A suffix tree is a data structure that can be built in $\theta(n)$ time. The power of suffix trees lies in quickly finding a particular substring of the string. It provides a fast implementation of many important string operations.

LEMMA 1. *Let $S$ be any given string of length $l$ over $(\Sigma, P)$ with count vector denoted by $\{Y_1, Y_2, \ldots, Y_k\}$ where each $Y_i \geq 0$ and $\sum_{i=1}^{k} Y_i = l$. Let $S'$ be any string which has $S$ as its prefix and is of length $l + l_1$. Then there exists some character $a_j \in \Sigma$ such that $X^2$ value of $S'$ is upper bounded by the $X^2$ value of the cover string $\lambda(S, a_j, l_1)$. The character $a_j$ is such that it has the maximum value of $\frac{2Y_j + l_1}{p_j}$ among all $j \in \{1, 2, \ldots, k\}$.*

PROOF. Let the $X^2$ values of strings $S$, $S'$ and $\lambda(S, a_j, l_1)$ be denoted by $X_S^2$, $X_{S'}^2$ and $X_\lambda^2$ respectively. We need to prove that $X_{S'}^2 \leq X_\lambda^2$.

By definition, the count vector of $\lambda(S, a_j, l_1)$ is $\{Y_1, Y_2, \ldots, Y_j + l_1, \ldots Y_k\}$. Further, let $Y_i'$ denote the frequency of character $a_i$ in $S'$ that are not present in $S$ (i.e., frequency of $a_i$ in the $l_1$ length suffix of $S'$). So, the count vector of $S'$ is $\{Y_1 + Y_1', Y_2 + Y_2' \ldots, Y_k + Y_k'\}$ where each $Y_m' \geq 0$ and $\sum_{i=1}^{k} Y_m' = l_1$. From the definition of $X^2$ statistic given in (5), we have

$$X_S^2 = \sum_{m=1}^{k} \frac{Y_m^2}{l p_m} - l \qquad (6)$$

$$X_\lambda^2 = \sum_{m=1, m \neq j}^{k} \frac{Y_m^2}{(l+l_1) p_m} + \frac{(Y_j + l_1)^2}{(l+l_1) p_j} - (l + l_1)$$

$$= \sum_{m=1}^{k} \frac{Y_m^2}{(l+l_1) p_m} + \frac{2 Y_j l_1 + l_1^2}{(l+l_1) p_j} - (l + l_1) \qquad (7)$$

$$X_{S'}^2 = \sum_{m=1}^{k} \frac{(Y_m + Y_m')^2}{(l+l_1) p_m} - (l + l_1)$$

$$= \sum_{m=1}^{k} \frac{Y_m^2}{(l+l_1) p_m} + \sum_{m=1}^{k} \frac{2 Y_m Y_m' + Y_m'^2}{(l+l_1) p_m} - (l + l_1) \qquad (8)$$

The character $a_j$ is chosen such that it maximizes the quantity $\frac{2Y_j + l_1}{p_j}$ over all possible alphabets. So for any other character $a_m$ where $m \in \{1, 2, \ldots, k\}$ we have

$$\frac{2 Y_m + Y_m'}{p_m} \leq \frac{2 Y_m + l_1}{p_m} \leq \frac{2 Y_j + l_1}{p_j} \qquad (9)$$

Multiplying (9) by $Y_m'$ and summing it over $m$ we get

$$\sum_{m=1}^{k} \frac{2 Y_m Y_m' + Y_m'^2}{p_m} \leq \sum_{m=1}^{k} Y_m' \frac{2 Y_j + l_1}{p_j} \leq \frac{2 Y_j l_1 + l_1^2}{p_j} \qquad (10)$$

From (7), (8) and (10) we have

$$X_{S'}^2 \leq \sum_{m=1}^{k} \frac{Y_m^2}{(l+l_1) p_m} + \frac{2 Y_j l_1 + l_1^2}{(l+l_1) p_j} - (l + l_1) = X_\lambda^2.$$

$\square$

The next lemma states that the $X^2$ value of a string can always be increased by adding a particular character to it.

LEMMA 2. *Let $S$ be any given string of length $l$ over $(\Sigma, P)$ with count vector denoted by $\{Y_1, Y_2, \ldots, Y_k\}$ where each $Y_i \geq 0$ and $\sum_{i=1}^{k} Y_i = l$. There always exists some character $a_j$ such that by appending it to $S$, the $X^2$ value of resultant string $S'$ becomes greater than that of $S$. The character $a_j$ is such that it has the maximum value of $\frac{Y_j}{p_j}$ among all $j \in \{1, 2, \ldots, k\}$.*

PROOF. Let the $X^2$ values of strings $S$ and $S'$ be denoted by $X_S^2$ and $X_{S'}^2$ respectively. We need to prove that $X_S^2 < X_{S'}^2$.

The string $S'$ is the resultant string obtained by appending alphabet $a_j$ to the string $S$, so the count vector of $S'$ is $\{Y_1, \ldots, Y_j + 1, \ldots, Y_k\}$. From (5), we have

$$X_S^2 = \sum_{m=1}^{k} \frac{Y_m^2}{lp_m} - l \quad (11)$$

$$X_{S'}^2 = \sum_{m=1}^{k} \frac{Y_m^2}{(l+1)p_m} + \frac{2Y_j + 1}{(l+1)p_j} - (l+1) \quad (12)$$

From (11) and (12) we have

$$X_{S'}^2 - X_S^2$$
$$= \sum_{m=1}^{k} \frac{Y_m^2}{(l+1)p_m} + \frac{2Y_j + 1}{(l+1)p_j} - (l+1) - \sum_{m=1}^{k} \frac{Y_m^2}{lp_m} + l$$
$$= \frac{1}{l(l+1)} \left[ \frac{(2Y_j + 1)l}{p_j} - l(l+1) - \sum_{m=1}^{k} \frac{Y_m^2}{p_m} \right] \quad (13)$$

The character $a_j$ is chosen such that it maximizes $\frac{Y_j}{p_j}$ over all $j$. So we have

$$\frac{Y_m}{p_m} \leq \frac{Y_j}{p_j} \quad \forall m \in \{1, 2, \ldots, k\} \quad (14)$$

Multiplying (14) by $Y_m$ and summing it over $m$ we get

$$\sum_{m=1}^{k} \frac{Y_m^2}{p_m} \leq \frac{Y_j}{p_j} \sum_{m=1}^{k} Y_m = \frac{lY_j}{p_j} \quad (15)$$

Putting (15) into (13) we get

$$X_{S'}^2 - X_S^2 \geq \frac{1}{l(l+1)p_j} \left[ (2Y_j + 1)l - l(l+1)p_j - lY_j \right]$$
$$= \frac{1}{l(l+1)p_j} \left[ l(Y_j - lp_j) + l(1 - p_j) \right] \quad (16)$$

Again, from (14) we have:

$$Y_m p_j \leq Y_j p_m \Rightarrow p_j \sum_{m=1}^{k} Y_m \leq Y_j \sum_{m=1}^{k} p_m \Rightarrow lp_j \leq Y_j \quad (17)$$

Putting (17) into (16) and using $p_j < 1$, we get

$$X_{S'}^2 - X_S^2 > 0.$$

□

In the next result, we show that the $X^2$ value of any string $S'$ having $S$ as its prefix is upper bounded by $X^2$ value of the chain cover of $S$.

THEOREM 1. *Let $S$ be any given string of length $l$ over $(\Sigma, P)$ with count vector denoted by $\{Y_1, Y_2, \ldots, Y_k\}$ where each $Y_i \geq 0$ and $\sum_{i=1}^{k} Y_i = l$. Further, let $S'$ be any string which has $S$ as its prefix and is of length less than or equal to $l + l_1$. Then there exists some character $a_j \in \Sigma$ such that $X^2$ value of $S'$ is less than $\lambda(S, a_j, l_1)$. The character $a_j$ is such that it has the maximum value of $\frac{2Y_j + l_1}{p_j}$ among all $j \in \{1, 2 \ldots k\}$.*

PROOF. The proof follows directly from the results stated in Lemma 1 and Lemma 2. From Lemma 2, we can say that there always exists a character such that appending it increases the $X^2$ value of $S'$. Hence, we keep appending the string $S'$ with such characters till its length becomes $l + l_1$. We call the resultant string $S_c$. Clearly, $S_c$ has $S$ as its prefix and is of length $l + l_1$ and $X^2$ value of $S'$ is less than or equal to $X^2$ value of $S_c$. The character $a_j$ is such that maximizes $\frac{2Y_j + l_1}{p_j}$ over all $j \in \{1, 2, \ldots, k\}$; so using Lemma 1, we can say that the $X^2$ value of $S_c$ is less than the $X^2$ value of $\lambda(S, a_j, l_1)$. This further implies that $X^2$ value of $S'$ is less than or equal to the $X^2$ value of $\lambda(S, a_j, l_1)$. □

We next formally describe our algorithm for finding the most significant substring (MSS).

## 4. THE MSS ALGORITHM

The algorithm looks for the possible candidates of MSS in an ordered fashion. The pseudocode is shown in Algorithm 1. The loop in line 2 iterates over the start positions of the substrings while the loop in line 3 iterates over all the possible lengths of the substrings from a particular start position. We keep track of the maximum $X^2$ value of any substring computed by our algorithm by storing it in a variable $X_{max}^2$. For a given substring $S[i \ldots l']$, we calculate its $X^2$ value, which is stored in $X_l^2$ (line 5). If $X_l^2$ turns out to be greater than $X_{max}^2$ then $X_{max}^2$ is updated accordingly (line 7).

The character $a_t$ is chosen such that it maximizes the value of $\frac{2Y_j + x}{p_j}$ over all $j$ (line 9 of the pseudocode). This property is necessary for the application of the result stated in Theorem 1. Denoting the $X^2$ value of a chain cover of $S[i \ldots l']$ over $x$ symbols of character $a_t$ by $X_\lambda^2$, the result stated in Theorem 1 states that the $X^2$ value of any substring of the form $S[i \ldots (l' + m)]$ for $m \in \{0, 1, \ldots, x\}$ is upper bounded by $X_\lambda^2$. We choose $x$ such that it is maximized within the constraint that $X_\lambda^2$ is guaranteed to be less than or equal to $X_{max}^2$. Then, under the given constraint, we can skip checking all substrings of the form $S[i \ldots (l' + m)]$ for $m \in \{0, 1, \ldots, x\}$ as their $X^2$ values are not greater than $X_{max}^2$. So, we directly increment $l$ by $x$ (line 14). Next, we find out what the ideal choice of $x$ is.

We denote the count vector of substring $S[i \ldots l']$ of length $l$ by $\{Y_1, Y_2, \ldots, Y_k\}$. The count vector of cover chain is given by $\{Y_1, Y_2 \ldots, Y_t + x, \ldots, Y_k\}$ where $Y_t$ denotes the frequency of

---

**Algorithm 1** Algorithm for finding the most significant substring (MSS)

1: $X_{max}^2 \leftarrow 0$
2: **for** $i = n$ to $1$ **do**
3:   **for** $l = 0$ to $n - i$ **do**
4:     $l' \leftarrow i + l$
5:     $X_l^2 \leftarrow X^2$ value of $S[i \ldots l']$
6:     **if** $X_l^2 > X_{max}^2$ **then**
7:       $X_{max}^2 \leftarrow X_l^2$
8:     **end if**
9:     $t \leftarrow m$ s.t. $\forall m \in \{1, 2, \ldots, k\}, \frac{2Y_m + x}{p_m}$ is maximum
10:     $a \leftarrow 1 - p_t$
11:     $b \leftarrow 2Y_t - 2lp_t - p_t X_{max}^2$
12:     $c \leftarrow (X_l^2 - X_{max}^2)lp_t$
13:     $x \leftarrow \lceil \frac{-b + \sqrt{b^2 - 4ac}}{2a} \rceil$
14:     Increment $l$ by $x$
15:   **end for**
16:   $i \leftarrow i - 1$
17: **end for**
18: **return** $X_{max}^2$

character $a$ in the algorithm. By definition of $X^2$ from (5),

$$X_l^2 = \sum_{m=1}^{k} \frac{(Y_m)^2}{lp_m} - l \qquad (18)$$

and

$$X_\lambda^2 = \sum_{m=1}^{k} \frac{(Y_m)^2}{(l+x)p_m} + \frac{2xY_t + x^2}{(l+x)p_t} - (l+x)$$

$$= \frac{l(X_l^2 + l)}{(l+x)} + \frac{2xY_t + x^2}{(l+x)p_t} - (l+x) \qquad (19)$$

We want to maximize $x$ with the constraint that $X_\lambda^2 \leq X_{max}^2$. From (19) we have,

$$\frac{l(X_l^2 + l)}{(l+x)} + \frac{2xY_t + x^2}{(l+x)p_t} - (l+x) \leq X_{max}^2 \qquad (20)$$

On multiplying (20) by $(l+x)p_t$ and rearranging, the constraint simplifies to

$$(1-p_t)x^2 + (2Y_t - 2lp_t - p_t X_{max}^2)x + (X_l^2 - X_{max}^2)lp_t \leq 0 \qquad (21)$$

Eq. (21) is a quadratic equation in $x$ with $a = 1 - p_t > 0$, $b = 2Y_t - 2lp_t - p_t X_{max}^2$ and $c = (X_l^2 - X_{max}^2)lp_t \leq 0$ ($X_l^2 \leq X_{max}^2$). We need to maximize $x$ with the constraint that $ax^2 + bx + c \leq 0$. Thus, we choose $x$ as the positive root of the quadratic equation:

$$x = \frac{-b + \sqrt{b^2 - 4ac}}{2a} \qquad (22)$$

Since $a > 0$ and $c \leq 0$ we have $x \geq 0$. Further, since $x$ has to be an integer we choose $x$ as the greatest integer greater than or equal to the above value (line 13 of the algorithm).

## 5. ANALYSIS OF THE MSS ALGORITHM

We first show that the running time of the algorithm on an input string generated from a memoryless Bernoulli model is $O(kn^{3/2})$ with high probability where $n$ and $k$ denote the string and alphabet size respectively. For a string not generated from the null model, we will argue that the time taken by our algorithm on that string is less than the time taken by our algorithm on an equivalent string of the same size generated from the null model. Hence, the time complexity of our algorithm for any input string is $O(kn^{3/2})$ with high probability.

Let $S$ be any string drawn from a memoryless Bernoulli model. Let $T_{ij}$ denote the random variable that takes value 1 if $a_i$ occurs at position $S[j]$ and 0 otherwise. Each character of the string $S$ is independently drawn from a fixed probability distribution $P$, so the probability that $T_{ij} = 1$ is $p_i$. The frequency of character $a_i$ in the string $S$ denoted by the random variable $Y_i$ is the sum of $n$ Bernoulli random variables $T_{ij}$ where $j$ ranges from 1 to $n$. Since $Y_i$ is the sum of $n$ i.i.d. (independent and identically distributed) Bernoulli random variables, each having a success probability $p_i$, $Y_i$ follows a binomial distribution with parameters $n$ and $p_i$.

$$T_{ij} \sim Bernoulli(p_i)$$

$$Y_i = \sum_{j=1}^{n} T_{ij} \implies Y_i \sim Binomial(n, p_i) \qquad (23)$$

We state the following two standard results from the domain of probability distributions.

THEOREM 2. *For large values of $n$, the Binomial(n,p) distribution converges to Normal($\mu, \sigma^2$) distribution with the same mean and variance, i.e., $\mu = np$ and $\sigma^2 = np(1-p)$.*

PROOF. The proof uses the result of Central Limit Theorem. Please refer to [4] for the detailed proof.[3] □

It has been shown in [1] that for both $n$ and $np$ greater than a constant[4], the binomial distribution can be approximated by the normal distribution. Since all the probabilities $p_i$ in our setting are fixed, we can always find a constant (say $c$) such that for all $n$ greater than $c$, every $X_i \sim N(np_i, np_i(1-p_i))$ distribution. We use the following result to obtain the distribution of the $X^2$ statistic of any substring from a string generated using the null model.

THEOREM 3. *Let the random variable $Y_i$, $i \in \{1, 2 \ldots k\}$ follows $N(np_i, np_i(1-p_i))$ distribution with $\sum_{i=1}^{k} p_i = 1$ and the additional constraint that $\sum_{i=1}^{k} Y_i = n$. The random variable*

$$X^2 = \sum_{i=1}^{k} \frac{(Y_i - np_i)^2}{np_i} \qquad (24)$$

*then follows the chi-square distribution with $(k-1)$ degrees of freedom, denoted by $\chi^2(k-1)$.*

PROOF. It has to be noted that all $Y_i$'s in the theorem are not independent but have an added constraint that $\sum_{i=1}^{k} Y_i = n$. This is precisely the reason why the degrees of freedom of chi-square distribution is $k-1$ instead of $k$. A well known result is that the sum of squares of $n$ independent standard normal random variables follows a $\chi^2(k)$ distribution. The proof (which is slightly complicated) follows directly from this well known result. Please refer to [20] for the detailed proof. □

We will next prove that with high probability, the $X^2$ value of the MSS of $S$ generated using the null model is greater than $\ln n$. However, before that, we prove another useful result using elementary probability theory.

LEMMA 3. *Let $Z_{max}$ denote the maximum of $m$ i.i.d. random variables following $\chi^2(k)$ distribution. Then with probability at least $1 - O(1/m^2)$, for sufficiently large $m$ and for any constant $c > 0$, $\ln cm \leq Z_{max}$.*

PROOF. We first show this for $k = 2$. Let $f(x)$ and $F(x)$ denote the pdf and cdf of $\chi^2(2)$ distribution:

$$f(x; 2) = \frac{1}{2}e^{-x/2} \qquad F(x; 2) = 1 - e^{-x/2} \qquad (25)$$

We have

$$Z_{max} = max\{Z_1, Z_2, \ldots, Z_m\} \forall i,\ Z_i \sim \chi^2(k) \qquad (26)$$

For any constant $c > 0$ we have:

$$Pr\{Z_{max} > \ln cm\} = Pr\{\exists i,\ \text{s.t.}\ Z_i > \ln cm\}$$
$$= 1 - Pr\{\forall i, Z_i \leq \ln cm\} = 1 - (Pr\{Z_i \leq \ln cm\})^m$$
$$= 1 - (1 - e^{-\frac{1}{2}\ln cm})^m = 1 - (1 - \frac{1}{\sqrt{cm}})^m$$
$$\geq 1 - e^{-\sqrt{m/c}} \geq 1 - O(1/m^2) \qquad (27)$$

In the above proof we only utilized the asymptotic behavior of pdf and cdf of the $\chi^2(k)$ distribution. Since for any general $k$,

---

[3] In the above approximation, we can think of the binomial distribution as the discrete version of the normal distribution having the same mean and variance. So we do not need to account for the approximation error using the Berry-Esseen theorem [8].

[4] In general, the value of this constant is taken as 5 [1].

the asymptotic behavior of pdf and cdf of $\chi^2(k)$ distribution has the same dominating term $e^{-x/2}$, the above result is valid for any given $k$.[5]  □

LEMMA 4. *In the MSS algorithm, at any iteration in the loop over $i$, $X^2_{max} > \ln n'$ with probability at least $1 - O(1/n'^2)$ where $n' = n - i$.*

PROOF. We can verify from the pseudo code (Algorithm 1) that before we begin the loop in line 2 for $i = i_0$, we have checked all the substrings that are potential candidates for MSS of $S$ starting at $i > i_0$. So, at this instance, the variable $X^2_{max}$ stores the maximum $X^2$ value of any substring of the string $S[(i_0+1)\ldots n]$. In other words, the variable $X^2_{max}$ would store the maximum of $^{n'}C_2 = O(n'^2)$ (where $n' = n - i_0$) random variables each following the same $\chi^2(k-1)$ distribution. However, since these $O(n'^2)$ substrings are not mutually independent, the result of Lemma 3 cannot be directly applied in this case.

However, we can still say that a subset of at least $O(n')$ substrings are independent, with each substring following a $\chi^2(k-1)$ distribution. One way of constructing a mutually independent subset of size $O(n')$ is by choosing $n'/c$ substrings each of length $c$ such that they do not share any character among them, i.e., the $i^{th}$ substring in this set is $S[(ci-c)\ldots(ci-1))]$ where $c$ is a constant such that the binomial distribution can be approximated by the normal distribution for all strings of length greater than or equal to $c$. Since all characters of the string $S$ are drawn independently from a fixed probability distribution, all the substrings in the subset are mutually independent, and since length of all these substrings are greater than $c$, $X^2$ statistics of these substrings follow the $\chi^2(k-1)$ distribution. Consequently, the value of $X^2_{max}$ in our algorithm is greater than the max of at least $O(n')$ $\chi^2(k-1)$ i.i.d. random variables. Putting the value of $m = n'/c$ in the result of Lemma 3, we can prove the above result.  □

LEMMA 5. *On an input string generated from the null model, with high probability ($> 1 - \epsilon$ for any constant $\epsilon > 0$) the number of substrings skipped (denoted by $x$) in any iteration of the loop on $l$ in the MSS algorithm is $\omega(\sqrt{l})$ for sufficiently large values of $l$. Hence, $\epsilon$ can be set so close to 0 that with probability practically equal to 1, the number of substrings skipped $x$ in any iteration is at least $\omega(\sqrt{l})$.*

PROOF. As stated in (22), the number of substrings skipped in any iteration of the loop on $l$ is

$$x = \frac{-b + \sqrt{b^2 - 4ac}}{2a} \quad (28)$$

We will prove that in the string generated from the null model, with high probability $b \leq \frac{1}{2}\sqrt{lp_t \ln l}$ and $c \geq -\frac{1}{2} lp_t \ln l$. These bounds help us in guaranteeing that $x = \omega(\sqrt{l})$ with high probability. In order to prove the bounds on $b$ and $c$, we first prove that the following conditions hold with high probability.

(i) From the result stated in Lemma 4, for any constant $\epsilon_1 > 0$, we have with probability at least $1 - O(1/n'^2) > 1 - \epsilon_1$ that $X^2_{max} > \ln n'$ where $n' = n - i$. In the algorithm, $l$ in the loop iterates from 0 to $n - i$, so we have $l \leq n'$. Hence, $X^2_{max} > \ln l$ with probability at least $1 - \epsilon_1$.

(ii) Suppose $Y_t$ denotes the frequency of alphabet $a_t$ in the string $S[i\ldots l']$ of length $l$. As denoted in (23) it is the sum of $l$ independent Bernoulli random variables $T_{ij}$ each with expectation $p_t$; so, $E[Y_t] = lp_t$. Also, we have $Pr(T_{ij} \in [0,1]) = 1$. Now, using the Hoeffding's inequality [16], we get

$$Pr\{Y_t - E[Y_t] < t\} \geq 1 - e^{-\frac{2t^2 l^2}{\sum_{i=1}^{l}(b_i - a_i)^2}} \quad (29)$$

---
[5]The term of $x^{k/2-1}e^{-x/2}$ occurring in pdf of a general $k$ is asymptotically less than $e^{-x/2+\epsilon}$ and greater than $e^{-x/2-\epsilon}$ for any $\epsilon > 0$, which is independent of $k$.

Substituting $E[Y_t] = lp_t$, $t = \frac{1}{4}\sqrt{lp_t \ln l}$, $a_i = 0$ and $b_i = 1$, we have for any constant $\epsilon_2 > 0$

$$Pr\{Y_t - lp_t < \frac{1}{4}\sqrt{lp_t \ln l}\} \geq 1 - e^{-\frac{2lp_t \ln l}{16l}}$$

$$= 1 - l^{-\frac{p_t}{8}} \geq 1 - \epsilon_2 \quad (30)$$

(iii) As stated in Theorem 3, the $X^2$ value of substring $S[i\ldots l']$ of length $l$ denoted by $X^2_l$ follows the $\chi^2$ distribution. Further, using the definition of cdf of $\chi^2$ distribution denoted by $F_x$, we have for any constant $\epsilon_3 > 0$

$$Pr\{X^2_l < \frac{\ln l}{2}\} = F_x(\frac{\ln l}{2}) = 1 - e^{-\frac{\ln l}{4}} \geq 1 - \epsilon_3 \quad (31)$$

We choose constants $\epsilon_1$, $\epsilon_2$ and $\epsilon_3$ small enough such that for any constant $\epsilon > 0$, $1 - \epsilon_1 - \epsilon_2 - \epsilon_3 > 1 - \epsilon$. Thus combining the above three conditions, the following results hold with probability $1 - \epsilon$:

$$b = 2(Y_t - lp_t) - p_t X^2_{max} \leq 2(Y_t - lp_t) \leq \frac{1}{2}\sqrt{lp_t \ln l} \quad (32)$$

$$c = lp_t(X^2_l - X^2_{max}) \leq lp_t(\frac{1}{2}\ln l - \ln l) \leq -\frac{1}{2} lp_t \ln l \quad (33)$$

$$a = 1 - p_t \leq 1 \quad (34)$$

We use the fact that if any positive $x$ satisfies the equation $a'x^2 + b'x + c' \leq 0$ then it also satisfies the equation $ax^2 + bx + c \leq 0$ if $a \leq a'$, $b \leq b'$ and $c \leq c'$. So substituting upper bounds of $a$, $b$ and $c$ in (28) and maximizing $x$ in (28) we have with probability $1 - \epsilon$

$$x \geq \frac{1}{2}(\sqrt{\frac{1}{4}lp_t \ln l + 2lp_t \ln l} - \frac{1}{2}\sqrt{lp_t \ln l})$$

$$= \frac{1}{2}(\sqrt{\frac{9}{4}lp_t \ln l} - \frac{1}{2}\sqrt{lp_t \ln l})$$

$$= \frac{1}{2}\sqrt{lp_t \ln l} = \Omega(\sqrt{l \ln l}) = \omega(\sqrt{l}) \quad (35)$$

□

Further, in Algorithm 1, except line 9, all the steps inside the loop over $l$ in line 3 can be performed in constant time. However, if we can determine the frequencies of all of the characters in the substring $S[i\ldots l']$ in $O(1)$ time, then we can find the character $a_t$ (line 9) in $O(k)$ time. For this purpose, we maintain one count array for each character $a_t$, $\forall t = 1, \ldots, k$, where the $i^{th}$ element of the count array stores the number of occurrence of $a_t$ up to the $i^{th}$ position in the string. Each count array can be preprocessed in $O(n)$ time. Consequently, each iteration of the loop over $l$ in line 3 takes $O(k)$ time. Further the loop over $i$ in line 2 iterates $n$ times. Now, we only need to compute the number of iterations of the loop over $l$ for which we use the next lemma.

LEMMA 6. *The expected number of iterations of the loop on $l$ (in line 3 of the MSS algorithm) for each value of $i$ is $O(\sqrt{n})$.*

**Algorithm 2** Algorithm for finding the *top-t* substrings
1: $T \leftarrow$ Min Heap on $t$ elements all initialized to 0
2: **for** $i = n$ to 1 **do**
3:    **for** $l = 0$ to $n - i$ **do**
4:       $l' \leftarrow i + l$
5:       $X^2_{max\_t} \leftarrow$ Find Min(T)
6:       $X^2_l \leftarrow X^2$ value of $S[i \ldots l']$
7:       **if** $X^2_l > X^2_{max\_t}$ **then**
8:          Extract Min($T$)
9:          Insert $X^2_l$ in $T$
10:      **end if**
11:      $t \leftarrow m$ s.t. $\forall m \in \{1, 2, \ldots, k\}$, $\frac{2Y_m + x}{p_m}$ is maximum
12:      $a \leftarrow 1 - p_t$
13:      $b \leftarrow 2Y_t - 2lp_t - p_t X^2_{max\_t}$
14:      $c \leftarrow (X^2_l - X^2_{max\_t})lp_t$
15:      $x \leftarrow \left\lceil \frac{-b + \sqrt{b^2 - 4ac}}{2a} \right\rceil$
16:      Increment $l$ by $x$
17:    **end for**
18:    $i \leftarrow i - 1$
19: **end for**
20: **return** $T$

**Algorithm 3** Algorithm for finding all substrings having $X^2$ value greater than $\alpha_0$
1: $S_{\alpha_0} \leftarrow \phi$
2: **for** $i = n$ to 1 **do**
3:    **for** $l = 0$ to $n - i$ **do**
4:       $l' \leftarrow i + l$
5:       $X^2_l \leftarrow X^2$ value of $S[i \ldots l']$
6:       **if** $X^2_l > \alpha_0$ **then**
7:          $S_{\alpha_0} \leftarrow S_{\alpha_0} \cup S[i \ldots l']$
8:      **end if**
9:      $t \leftarrow m$ s.t. $\forall m \in \{1, 2, \ldots, k\}$, $\frac{2Y_m + x}{p_m}$ is maximum
10:      $a \leftarrow 1 - p_t$
11:      $b \leftarrow 2Y_t - 2lp_t - p_t\alpha_0$
12:      $c \leftarrow (X^2_l - \alpha_0)lp_t$
13:      $x \leftarrow \max \left\{ \left\lceil \frac{-b + \sqrt{b^2 - 4ac}}{2a} \right\rceil, 1 \right\}$
14:      Increment $l$ by $x$
15:    **end for**
16:    $i \leftarrow i - 1$
17: **end for**
18: **return** $S_{\alpha_0}$

PROOF. Let $T(r)$ be the number of iterations of the loop over $l$ required for $l$ to reach $r$. We have shown in Lemma 5 that in each iteration, the number of substrings skipped $x$ is $\omega(\sqrt{l})$. Thus, $l$ in the next iteration will reach from $r$ to $r + \omega(\sqrt{r})$. This gives us the following recursive relation:

$$T(r + c\sqrt{r}) \leq T(r) + O(1) = T(r) + q \qquad (36)$$

It can be shown that the solution to the above relation is $O(\sqrt{n})$. Please refer to Lemma 7 in the appendix for detailed proof. □

Since each iteration of the loop over $l$ in line 3 takes $O(\sqrt{n})$ time, the time taken by the algorithm on an input string generated by the null model is $O(kn^{3/2})$ which is $O(n^{3/2})$ since $k$ is taken as a constant in our problem setting. Thus, we have shown that the running time of the algorithm on an input string generated from a memoryless Bernoulli model is $O(kn^{3/2})$ with high probability.

### 5.1 Nature of the String

As it can be verified from the definition, the $X^2$ value of a substring increases when the expected and observed frequencies begin to diverge. Thus, the individual substrings of a string not generated from the null model are expected to have higher $X^2$ values which, in turn, increases the $X^2_{max}$. Further, it can be verified from (22) that the number of substrings skipped, $x$, increases on increasing $X^2_{max}$ as we have to maximize $x$ such that the constraint $X^2_\lambda \leq X^2_{max}$ is satisfied. If $X^2_{max}$ is large, it gives a larger window for $X^2_\lambda$ which allows the choice of a larger $x$. Hence, the time taken by our algorithm on an input string not generated from null model is less than the time taken by our algorithm on an equivalent string of the same size generated from the null model. So, the time complexity of our algorithm remains $O(n^{3/2})$ and is independent of the nature of the input string. Section 7.1.2 gives the details on how our algorithms perform on different types of strings.

## 6. OTHER VARIANTS OF THE PROBLEM

### 6.1 Top-t Substrings

The algorithm for finding the *top-t* statistically significant substrings (Algorithm 2) is same as the algorithm for finding the MSS except that $X^2_{max\_t}$ stores the $t^{th}$ largest $X^2$ value among all substrings seen till that particular instant by the algorithm. We maintain a min-heap $T$ of size $t$ for storing the *top-t* $X^2$ values seen by the algorithm. The heap $T$ is initially empty and $X^2_{max\_t}$ always stores the top (minimum) element of the heap. If $X^2_l$ is computed to be greater than $X^2_{max}$, then we extract the minimum element of $T$ (which now no more is a part of *top-t* substrings) and insert the new $X^2_l$ value into the heap. Now, $X^2_{max\_t}$ points to the new minimum of the heap. Finally, at the end of the algorithm we return the heap $T$ which contains the *top-t* $X^2$ values among all the substrings of string $S$.

The analysis of this algorithm is same as the algorithm for MSS except that we now need to show that $X^2_{max\_t}$ is greater than $\ln n$ with probability greater than any constant. This still holds true for any $t < \omega(n)$ (please refer to Lemma 8 in the appendix for detailed proof). Moreover, inside the for loop on $l$, we now perform insertion and extract-min operations on a heap $T$ of size $t$; so each iteration of the loop over $l$ now requires $O(k + \log t)$ time. Thus, the total time complexity of the algorithm for finding the top $t$ substrings is $O((k + \log t)n^{3/2})$ for $t < \omega(n)$.

### 6.2 Significance Greater Than a Threshold

The algorithm for finding all substrings having $X^2$ value greater than a threshold $\alpha_0$ (Algorithm 3) is again essentially the same as the MSS algorithm except that the $X^2_{max}$ constantly remains $\alpha_0$ at every iteration. We maintain $S_{\alpha_0}$ as a set of all substrings having $X^2$ value greater than $\alpha_0$. We skip all substrings that cannot be a part of $S_{\alpha_0}$, i.e., whose cover strings have $X^2$ value not greater than $\alpha_0$.

Next, we analyze the time complexity of the algorithm on varying $\alpha_0$. We again revert to (22):

$$x = \max \left\{ \left\lceil \frac{-b + \sqrt{b^2 - 4ac}}{2a} \right\rceil, 1 \right\} \qquad (37)$$

where $a = 1 - p_t > 0$, $b = 2Y_t - 2lp_t - p_t\alpha_0$ and $c = (X^2_l - \alpha_0)lp_t \leq 0$. If $\alpha_0 < X^2_l$ then $c$ in the above equation is positive. Consequently, as $x$ takes the value 1, the number of iterations of the loop on $l$ is $O(n)$. Hence, the time complexity of the algorithm is $O(kn^2)$. However, the time complexity decreases sharply on

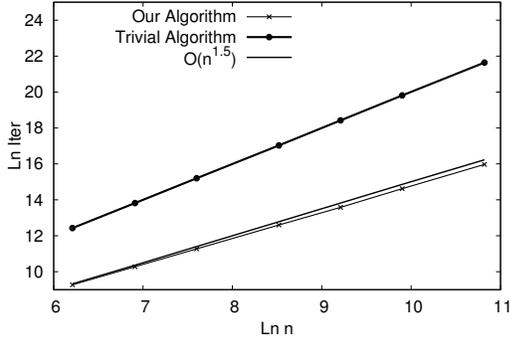

(a) Number of iterations with string length $n$ ($k=2$).

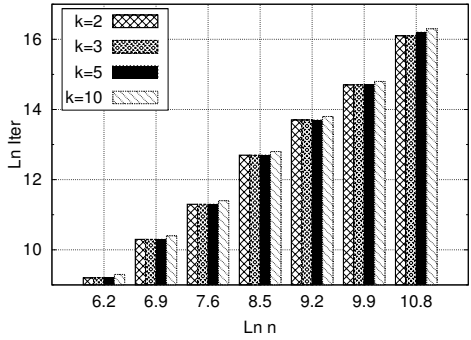

(b) Number of iterations with alphabet size $k$.

Figure 1: Analysis of time complexity for finding the MSS.

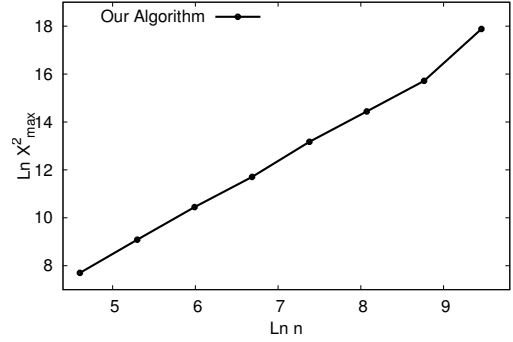

Figure 2: Variation of $X^2_{max}$ with string length $n$ ($k=2$).

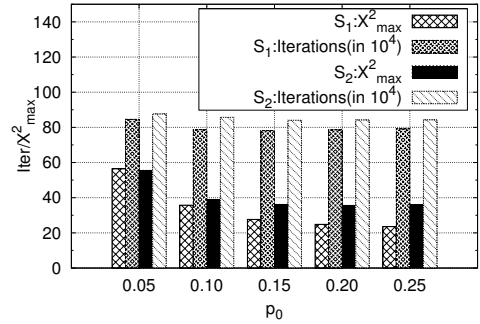

Figure 3: $X^2_{max}$ and number of iterations for different multinomial strings. $S_1 : n = 10^4, k = 3, P = \{p_0, 0.5 - p_0, 0.5\}$; $S_2 : n = 10^4, k = 5, P = \{p_0, 0.5 - p_0, 0.1, 0.2, 0.2\}$.

increasing $\alpha_0$. Once $\alpha_0$ becomes sufficiently greater than $X^2_l$, the term $c \approx -\alpha_0 l p_t$ starts predominating $b$, and $x$ in each step is effectively $c/a$ which is $O(\sqrt{\alpha l})$.[6] Hence, the recurrence relation of the number of iterations of the loop on $l$ in this case is

$$T(l + O(\sqrt{\alpha_0 l})) = T(l) + 1 \qquad (38)$$

It can be again shown with the help of Lemma 7 in the appendix that the solution to the recursive relation is $O(\sqrt{l/\alpha_0})$. So the total time complexity of the algorithm is $O(kn\sqrt{n/\alpha_0})$.

## 6.3 MSS Greater Than a Given Length

The algorithm for finding the most significant substring among all substrings having length greater than a given length $\Gamma_0$ is exactly the same as the MSS algorithm except that now we ignore any substring whose length is not greater than $\Gamma_0$. This means the loop on $l$ starts with $\Gamma_0$ instead of 0 and loop on $i$ goes on till $n - \Gamma_0$ instead of $n$. The time complexity of the algorithm decreases not just because of less number of substrings evaluated in this case but also because the skip $x$ in our algorithm is a function of $l$ and it increases with increasing values of $l$. Hence, the recursive relation for the loop over $l$ in this case is the same with only the base case different: $T(\Gamma_0) = 1$ instead of $T(1) = 1$. The solution to this recurrence relation is $O(\sqrt{n} - \sqrt{\Gamma_0})$. Since there are $n - \Gamma_0$ iterations of loop in $i$, the total time complexity of the algorithm is $O(k(n - \Gamma_0)(\sqrt{n} - \sqrt{\Gamma_0}))$ which is effectively $O(kn^{3/2})$.

---

[6]In a substring generated from a memoryless Bernoulli distribution, $X^2$ follows a $\chi^2$ distribution with constant mean and variance. Hence, it can be shown with high probability that $X^2_l$ is a small constant.

## 7. EXPERIMENTAL ANALYSES AND APPLICATIONS

The experimental results shown in this section are for C codes run on Macintosh platform on a machine with 2.3 GHz Intel dual core processor and 4 GB, 1333 MHz RAM. Each character of a synthetic string was generated independently from the underlying distribution assumed using the standard uniform $(0, 1)$ random number generator in C.

### 7.1 Synthetic Datasets

#### 7.1.1 Time Complexity of Finding MSS

The first experiment is on the time complexity of our algorithm for finding the most significant substring. Figure 1a depicts the comparison of number of iterations required by our algorithm vis-à-vis the trivial algorithm for input strings of different lengths ($n$) generated from the null model for an alphabet of size 2. The number of iterations of our algorithm when plotted on a logarithmic scale increases linearly with the logarithm of the string size with a slope close to 1.5. Hence, we can claim that the empirical time complexity of our algorithm for an input string generated by null model is also $O(n^{1.5})$.

The effect of varying the alphabet size is shown in Figure 1b for different string lengths. It can be observed that, as expected, varying the alphabet size has no significant effect on the number of iterations of the algorithm.

Figure 2 shows that the expected $X^2_{max}$ increases linearly with $n$ with slope $\sim 2$ which supports our claim in Lemma 4 that for sufficiently large $n$, $X^2_{max}$ is greater than $\ln n$ with high probability.

Finally, Figure 3 plots the variation of $X^2_{max}$ and iterations of the loop over $l$ for different heterogeneous multinomial distributions

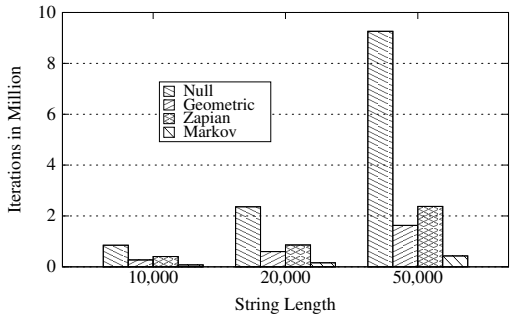

(a) Varying $n$ ($k = 5$).

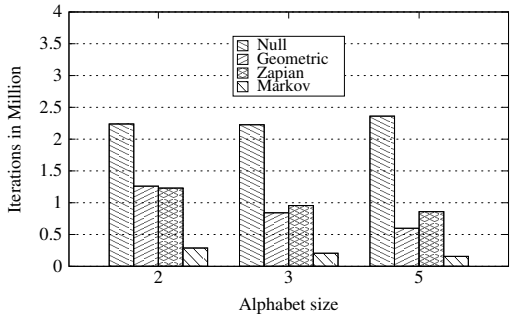

(b) Varying $k$ ($n = 20000$).

Figure 4: Comparison of time taken by our algorithm on strings not generated by the null model.

and different alphabet sizes. It is evident that change in the probability $p_0$ of occurrence of character $a_0$ only changes the $X^2_{max}$ but has no significant effect on the number of iterations taken by our algorithm. It can be intuitively seen that the change in $p_0$ is effectively canceled out by the change in $X^2_{max}$, so the number of characters skipped ($x$ in Eq. (22)) roughly remains the same.

### 7.1.2 Strings Not Generated Using the Null Model

We now investigate the results for input strings *not* generated from the null model in addition to an equivalent length input string generated from the null model which is a memoryless Bernoulli source where the multinomial probabilities of all the characters are *equal*. The different types of other strings that we compare are:

(a) *Geometric string:* A string generated from a memoryless multinomial Bernoulli source but the multinomial probabilities of all the characters are different. The probability of occurrence of a character decreases *geometrically*. Hence, the probability of occurrence of character $a_i$ is proportional to $1/2^i$.

(b) *Harmonic string:* A string generated from a memoryless multinomial Bernoulli source but the multinomial probabilities of all the characters are different. The probability of occurrence of a character decreases *harmonically*. Hence, the probability of occurrence of character $a_i$ is proportional to $1/i$.

(c) *Markov string:* A string generated by a Markov process, i.e., the occurrence of a character depends on the previous character. The state transition probability of character $a_j$ following character $a_i$ is proportional to $1/2^{(i-j) \bmod k}$.

The number of iterations for our algorithm on different values of string length ($n$) and alphabet size ($k$) are plotted in Figure 4. It can be verified that in all the cases, the string generated using the null model requires the maximum number of iterations which

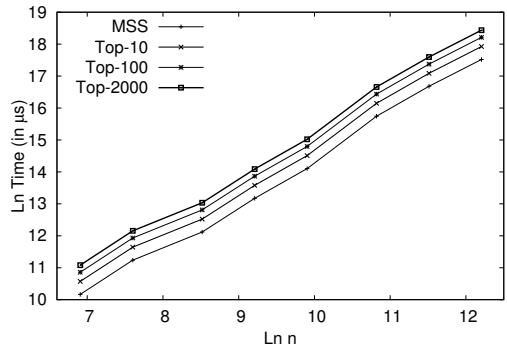

(a) Number of iterations with string length $n$.

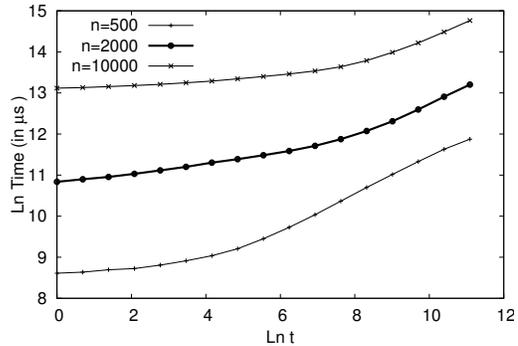

(b) Number of iterations with $t$.

Figure 5: Analysis of time complexity for finding the *top-t* set.

is in accordance with our theoretical claim in Section 5. The time taken by our algorithm on an input string not generated from a null model is upper bounded by the time taken on an equivalent size input string generated from the null model. This verifies that the time complexity of our algorithm is $O(kn^{3/2})$, independent of the type of the input string.

## 7.2 Other Variants

### 7.2.1 Top-t Significant Substrings

The time taken by the algorithm for finding the *top-t* set on varying string lengths for different values of $t$ is shown in Figure 5a. The linear increment in logarithmic scale with slope $\sim 1.5$ verifies that for any constant $t$ the time taken by our algorithm to find the top-t set is again $O((k + \log t)n^{1.5})$.

The time taken for different $t$ is shown in Figure 5b. The plot shows that till $t < \omega(n)$, the running time increases with slope 1.5, but once $t$ crosses the limit, the slope starts increasing towards 2. This is agreement with our theoretical analysis in Section 6.1.

### 7.2.2 Significance Greater Than a Threshold

Figure 6 depicts the number of iterations taken by the algorithms for finding all substrings greater than a threshold $\alpha_0$. As discussed in Section 6.2, the iterations decrease very sharply from $O(n^2)$ until $\alpha_0 = O(X^2_{max})$ after which it gradually decreases (as a function of $1/\sqrt{\alpha_0}$).

### 7.2.3 Substrings Greater Than a Given Length

The number of iterations taken by the algorithms for finding the MSS among all strings of length greater than $\Gamma_0$ is shown in Figure 7. As discussed in Section 6.3, the number of iterations slowly decreases as $\Gamma_0$ tends to $n$ before rapidly approaching 0.

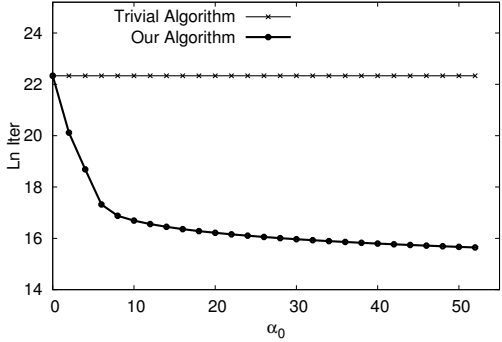

Figure 6: Number of iterations with $\alpha_0$ ($n = 10^5$, $k = 2$).

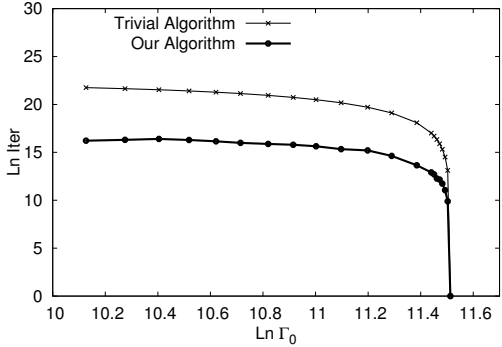

Figure 7: Number of iterations with $\Gamma_0$ ($n = 10^5$, $k = 2$).

## 7.3 Comparison with Existing Techniques

Table 1 presents the comparative results of our algorithm with the existing algorithms [13] for two different values of string size (averaged over different runs). As expected, results indicate that ARLM [13], being $O(n^2)$, does not scale well for larger strings, as opposed to our algorithm. AGMM [13], being $O(n)$ time, is very fast and outperforms all the algorithms in terms of time taken. However, being just a heuristic with no theoretical guarantee, it does not always lead to a solution that is close to the optimal. As can be verified from Table 1, the average $X^2_{max}$ string found by AGMM is significantly lower than the average $X^2_{max}$ value found by other algorithms. Further, since there are no guarantees on the lower bound of the $X^2_{max}$ value found by it relative to the optimal $X^2_{max}$ value, AGMM can lead to pretty bad solutions in some real datasets which are not as well behaved as the synthetic ones (Section 7.5). Finally, our algorithm requires only 3 seconds for a string as large as of length 80000 which signifies that for real life scenarios, the algorithm is practical.

## 7.4 Application in Cryptology

The correlation between adjacent symbols is of central importance in many cryptology applications [12]. The objective of a random number generator is to draw symbols from the null model. The independence of consecutive symbols is an important criterion for efficiency of a random number generator [12]. We define correlation between adjacent symbols in terms of the state transition probability. An ideal random binary string generator should generate the same symbol in next step with probability exactly 0.5. However, some random number generators which are inefficient might be biased towards generating the same symbol again with probability more than 0.5. Table 2 shows the comparison of $X^2_{max}$ for different lengths $n$ of string and different probabilities of generation of same symbol $p$ in the next iteration.

| Algo | String Size | Avg $X^2_{max}$ | Avg Time |
|---|---|---|---|
| Trivial | 20000 | 18.69 | 8.54s |
| Our | 20000 | 18.69 | 0.5s |
| ARLM | 20000 | 18.69 | 1.9s |
| AGMM | 20000 | 15.10 | 0.01s |
| Trivial | 80000 | 20.35 | 142.21s |
| Our | 80000 | 20.35 | 2.82s |
| ARLM | 80000 | 20.32 | 39.22s |
| AGMM | 80000 | 17.71 | 0.03s |

Table 1: Comparison with other techniques for synthetic datasets.

| $X^2_{max}$ | $p = 0.50$ | $p = 0.55$ | $p = 0.60$ | $p = 0.80$ |
|---|---|---|---|---|
| $n = 1000$ | 12.18 | 14.24 | 16.80 | 36.47 |
| $n = 5000$ | 15.12 | 17.67 | 21.52 | 48.79 |
| $n = 10000$ | 16.87 | 19.36 | 24.03 | 53.37 |
| $n = 20000$ | 17.89 | 21.48 | 25.70 | 60.61 |

Table 2: Variation of $X^2_{max}$ with $n$ and $p$.

It can be verified from the data that the $X^2_{max}$ is minimum for a string generated with $p = 0.5$ and increases with increasing $p$. Further, Figure 4 plots the variation of $X^2_{max}$ of a string generated using the null model with (logarithm of) the string length ($\ln n$). We observe a nice linear convergence with slope 2. This $X^2_{max}$ value can be used as a benchmark for a string of any length to measure the deviation from the null model. If the observed $X^2_{max}$ value of a string deviates significantly from the benchmark, it means that the string generated is not completely random but contains some kind of hidden correlation among the symbols. One of the major advantages of using the algorithm is in a scenario where only a substring of a string might deviate from the random behavior. Our algorithm will be able to capture such a substring without having to examine all the possible substrings[7].

## 7.5 Real Datasets

### 7.5.1 Analysis of Sports Data

The chi-square statistic can be used to find the best or worst career patches of sports teams or professionals. Boston Red Sox versus New York Yankees is one of the most famous and fiercest rivalries in professional sports [11]. They have competed against each other in over two thousand Major League Baseball games over a period of 100 years. Yankees have won 1132 (54.27%) of those games. However, we would like to analyze the time periods in which either of Yankees or Red Sox were particularly dominant against the other. The dominant periods should have large win ratio for a team over a sufficiently long stretch of games. If we encode the results in the form of a binary string whose letters denote a win or loss for a team, then these sufficiently long periods will contain results that significantly differ from the expected or average. Consequently, the $X^2$ value for the dominant periods will significantly differ from 0. We use the dataset obtained from www.baseball-reference.com.

The top five most significant patches found by our algorithm have been summarized in Table 3. The best period for Yankees was from mid 1920s to early 1930s in which they won more than 75% of the games. It was clearly the era of Yankees dominance in which they won 26 World Series championships and 39 pennants, compared to only 4 pennants for the Red Sox [11]. Alternatively, the best patch for Red Sox was a two-year period around 1912 in which they had close to 90% winning record; this is also referred to as the glory period in Red-Sox history [11].

---
[7]Such substrings will tend to exhibit large $X^2$ values and, hence, will be captured by our algorithm.

| Start | End | $X^2$ val | Games | Wins | Win% |
|---|---|---|---|---|---|
| 17-04-1924 | 06-06-1933 | 38.76 | 204 | 155 | 75.98% |
| 05-09-1911 | 01-09-1913 | 26.99 | 39 | 5 | 12.82% |
| 02-05-1902 | 27-07-1903 | 16.93 | 27 | 4 | 14.81% |
| 08-02-1972 | 28-07-1974 | 16.56 | 35 | 7 | 20.00% |
| 10-07-1960 | 07-09-1962 | 12.05 | 42 | 34 | 80.05% |

Table 3: Performance of Yankees against Red-Sox.

| Algorithm | $X^2$ val | Start | End | Time |
|---|---|---|---|---|
| Trivial | 38.76 | 17-04-1924 | 06-06-1933 | 0.142s |
| Our | 38.76 | 17-04-1924 | 06-06-1933 | 0.036s |
| ARLM | 38.76 | 17-04-1924 | 06-06-1933 | 0.032s |
| AGMM | 26.99 | 05-09-1911 | 01-09-1913 | 0.011s |

Table 4: Comparison with other techniques for sports data.

The comparative results of our algorithm with existing techniques are summarized in Table 4. As expected, our algorithm and AGMM finds the optimal solution but our algorithm outperforms the trivial algorithm and is almost as good as ARLM in terms of time (due to relatively small string size). Moreover, though AGMM is faster, it does not find the optimal solution. The best period found by AGMM was the second best (see Table 3) and has a significantly lower $X^2$ value.

### 7.5.2 Analysis of Stock Returns

Most financial models are based on the random walk hypothesis which is consistent with the efficient-market hypothesis [6]. They assume that the stock market prices evolve according to a random walk with a constant drift and, thus, the prices of the stock market cannot be predicted.[8]

We analyze the returns of three generic financial securities for which a long historical data is available. The Dow Jones Industrial Average is one of the oldest stock market index that shows the performance of 30 large publicly owned companies in the United States. Similarly, S&P 500 is another large capitalization-weighted index that captures the performance of 500 large-cap common stocks actively traded. Finally, the IBM common stock is representative of one of the oldest and largest publicly owned firms. We run the algorithms on the Dow Jones prices obtained since the year 1928 onwards (20906 days), S&P 500 since 1950 onwards (15600 days) and IBM since 1962 onwards (12517 days). The daily price data are obtained from finance.yahoo.com.

Given the randomness in the stock prices, we assume that the prices can increase (or decrease) each day with a fixed probability. The fixed probability is calculated as the ratio of days on which price went up (or down) to the total number of trading days. We find the statistically significant substrings of the binary string encoded with 1 for the day if the price of security went up and 0 otherwise. These substrings correspond to significantly long periods that contain a large ratio of days in which the stock price changed. The results are summarized in Table 5.

A lot of bad periods occurred during the Great Depression of 1930s, the recent dot-com bubble burst and mortgage recession periods of the last decade, whereas a number of good periods occurred during the economic boom of 1950s and 1960s. These observations verify that these statistically significant periods do not occur just due to randomness or chance alone, but are consequences of external factors as well. The identification of such significant patterns can help in identifying the relevant external factors. Finally, the $X^2$ values of these substrings can also be used in quantifying the historical risk of the securities which is one of the most important parameters that investment managers like to control.

---

[8]If the stock prices can be predicted then there is an arbitrage in the market which violates the efficient market hypothesis.

| Periods | Security | Start | End | Change |
|---|---|---|---|---|
| Good | Dow Jones | 24-02-1954 | 06-12-1955 | 68.10% |
| | Dow Jones | 25-06-1958 | 04-08-1959 | 43.52% |
| | S&P 500 | 15-09-1953 | 20-09-1955 | 97.07% |
| | S&P 500 | 09-12-1994 | 17-05-1995 | 17.92% |
| | IBM | 13-08-1970 | 06-10-1970 | 37.60% |
| | IBM | 26-10-1962 | 26-01-1968 | 252.0% |
| Bad | Dow Jones | 27-02-1931 | 04-05-1932 | -71.17% |
| | Dow Jones | 19-09-1929 | 14-11-1929 | -41.27% |
| | S&P 500 | 26-10-1973 | 21-11-1974 | -39.79% |
| | S&P 500 | 05-09-2000 | 12-03-2003 | -46.24% |
| | IBM | 31-03-2005 | 20-04-2005 | -21.20% |
| | IBM | 22-02-1973 | 13-08-1975 | -46.91% |

Table 5: Significant periods for the securities.

| Algo | Sec. | $X^2$ | Start | End | Change | Time |
|---|---|---|---|---|---|---|
| Trivial | Dow | 25.22 | 24-02-54 | 06-12-55 | 68.1% | 14.10s |
| Our | Dow | 25.22 | 24-02-54 | 06-12-55 | 68.1% | 0.89s |
| ARLM | Dow | 25.22 | 24-02-54 | 06-12-55 | 68.1% | 4.15s |
| AGMM | Dow | 19.53 | 24-01-66 | 09-04-85 | 325.0% | 0.03s |
| Trivial | S&P | 22.21 | 26-10-73 | 21-11-74 | -39.79% | 9.36s |
| Our | S&P | 22.21 | 26-10-73 | 21-11-74 | -39.79% | 0.63s |
| ARLM | S&P | 22.21 | 26-10-73 | 21-11-74 | -39.79% | 2.87s |
| AGMM | S&P | 13.44 | 22-04-66 | 09-05-66 | -6.44% | 0.03s |

Table 6: Comparison with other techniques for stock returns.

The comparative performance of our algorithm vis-à-vis the other techniques in finding the period with the highest $X^2$ value is summarized in Table 6. Again, as expected, our algorithm, trivial algorithm and ARLM find the same period for which the $X^2$ value is maximized. However, in this case, the time performance advantage of our algorithm over ARLM is pretty apparent. AGMM, though having the time advantage, does pretty badly in terms of identifying the maximum $X^2$ substring. Especially for S&P 500, it returns a substring that is not even close to the top few substrings.

## 8. CONCLUSIONS AND FUTURE WORK

In this paper, we chose to analyze the $X^2$ statistic in the context of a memoryless Bernoulli model. We experimentally saw that for a string drawn from such a model, the chi-square value of the most significant substring increases asymptotically as $(2 \ln n)$ where $n$ is the length of the string. However, the rigorous mathematical proof remains an interesting open problem. Such analysis of asymptotic behavior have significant applications in deciding the confidence interval with which the null hypothesis is rejected. Further, the analysis can be further extended to strings generated from Markov models, the most basic of which being the case when there is a correlation between adjacent characters.

The single dimensional problem of identification of the most significant substring can be extended to two-dimensional grid networks as well as general graphs. One potentially interesting application is in financial time series analysis of two securities that might not be very correlated in general, but might point to significant correlations during certain specific events such as recession. Such correlations are essential to most risk analysis techniques.

# APPENDIX

LEMMA 7. *The solution to the recursive relation* $T(\lceil l+c\sqrt{l}\rceil) \leq T(l)+1$ *with* $T(\alpha) = 1$ *for* $\alpha <= 1$ *where $l$ is a positive integer is* $O(\sqrt{l}/c)$. *More specifically,* $T(l) \leq \frac{4\sqrt{l}}{c} + c^2$.

PROOF. We prove this by induction. The base cases for $l < c^2$ are trivially satisfied. Further, for any positive integer $l \geq c^2$, assume that $T(l) \leq \frac{4\sqrt{l}}{c} + c^2$ is true for all positive integers $r$ such that $c^2 < r < \lceil l + c\sqrt{l} \rceil$. Hence,

$$T(\lceil l+c\sqrt{l}\rceil) \leq T(l) + 1$$
$$\leq \frac{4\sqrt{l}}{c} + c^2 + 1 = \sqrt{\left(\frac{4\sqrt{l}}{c}+1\right)^2} + c^2$$
$$= \frac{\sqrt{16l + 8c\sqrt{l} + c^2}}{c} + c^2$$
$$\leq \frac{\sqrt{16l + 9c\sqrt{l}}}{c} + c^2 \ [\because c^2 \leq l]$$
$$\leq \frac{4\sqrt{l+cl}}{c} + c^2 \qquad (39)$$

□

LEMMA 8. *In the algorithm for finding the* top-t *substrings, for any constant $\epsilon$ and $t < \omega(n)$, $X^2_{max\_t} < \ln n$ with probability at least $1-\epsilon$.*

PROOF. Let $Z_{max}$ denote the $t^{th}$ max of $m$ i.i.d. random variables following $\chi^2(k)$ distribution. As in the $X^2_{max}$ case in Algorithm 1, since asymptotic behavior of $\chi^2(k)$ distribution is same for all $k$, we again prove it only for $k = 2$, which is sufficient. Again, $f(x)$ and $F(x)$ denote the pdf and cdf of $\chi^2(2)$ distribution:

$$f(x;2) = \frac{1}{2}e^{-x/2} \qquad F(x;2) = 1 - e^{-x/2} \qquad (40)$$

We have

$$Z_{max\_t} = max\_t\{Z_1, Z_2, \ldots, Z_m\} \forall \text{ i}, Z_i \sim \chi^2(k) \qquad (41)$$

Now for each $Z_i$, we define a new Bernoulli random variable $Y_i$ which takes the value 1 if $Z_i > \ln m$ and 0 otherwise:

$$Pr\{Y_i = 1\} = Pr\{Z_i > \ln m\} = e^{-\frac{1}{2}\ln(m)} = \frac{1}{\sqrt{m}} \qquad (42)$$

Let $Y = \sum_{i=1}^{m} Y_i$; then $Y$ follows binomial distribution with probability of success $p = \frac{1}{\sqrt{m}}$. Further,

$$Pr\{Z_{max\_t} > \ln m\} = Pr\{Y \geq t\} \qquad (43)$$

Using the Chernoff's inequality for binomial distribution, for any constant $\epsilon > 0$,

$$Pr\{Y \geq t\} \geq 1 - e^{-\frac{(mp-t)^2}{2mp}} \geq 1 - e^{-\frac{(\sqrt{m}-t)^2}{2\sqrt{m}}} \qquad (44)$$

If $t < \omega(\sqrt{m})$, we can effectively ignore $t$ in the above equation. In that case, the above equation simplifies to

$$Pr\{Z_{max\_t} > \ln m\} \approx 1 - e^{-\frac{\sqrt{m}}{2}} \geq 1 - \epsilon \qquad (45)$$

Finally, again as in Algorithm 1, at least $O(n)$ substrings are independent. Therefore, the result holds. □